\documentclass[aps,onecolumn,11pt,floatfix,altaffilletter,tightenlines,showpacs,showkeys,notitlepage,nofootinbib]{revtex4-1}
\pdfoutput=1
\usepackage[colorlinks=true,citecolor=blue,linkcolor=blue]{hyperref}
\usepackage{url}
\usepackage[normalem]{ulem}
\usepackage{amsmath,amssymb}
\usepackage{epsfig}
\usepackage{graphicx}               
\usepackage{url}
\usepackage{color}
\usepackage{multirow}
\usepackage{placeins}
\usepackage[dvipsnames]{xcolor}
\usepackage{epstopdf}
\usepackage{comment}
\usepackage{soul}
\usepackage{tikz}
\usetikzlibrary{trees}
\usetikzlibrary{decorations.pathmorphing}
\usetikzlibrary{decorations.markings}

\definecolor{jblue}  {RGB}{20,50,100}
\definecolor{npurple}  {RGB} {153, 51, 204}
\definecolor{wred}   {RGB}{217,0,56}
\definecolor{white}   {RGB}{255,255,255}

\definecolor{korange}   {RGB}{235, 80,  43}
\definecolor{korange2}   {RGB}{245, 100,  63}
\definecolor{kyelloworange}   {RGB}{255, 210,  110}
\definecolor{kyelloworange2}   {RGB}{240, 170,  90}
\definecolor{kred}   {RGB}{204,  102, 153}
\definecolor{kpurple}   {RGB}{153,  61, 190}
\definecolor{kpurplelight}   {RGB}{213,  161, 230}

\usepackage[capitalise]{cleveref}
\usepackage{subfigure} 
\usepackage{lipsum}
\definecolor{red}{rgb}{1.0, 0, 0}
 \usepackage{gensymb}
 
\allowdisplaybreaks

\bibpunct{[}{]}{,}{n}{}{,}


\begin{document}

\title{Prospects for Finding Sterile Neutrino Dark Matter at KATRIN}

\author{Cristina Benso}   \email{cristina.benso@mpi-hd.mpg.de}
\author{Vedran Brdar}   \email{vedran.brdar@mpi-hd.mpg.de }
\author{Manfred Lindner} \email{manfred.lindner@mpi-hd.mpg.de}
\author{Werner Rodejohann} \email{werner.rodejohann@mpi-hd.mpg.de}
\affiliation{Max-Planck-Institut f\"ur Kernphysik, Saupfercheckweg 1, 
       69117~Heidelberg, Germany}

\begin{abstract}
\noindent 
We discuss under what circumstances a signal in upcoming laboratory searches for keV-scale sterile neutrinos would be compatible with those particles being a sizable part or all of dark matter. In the parameter space that will be experimentally accessible by KATRIN/TRISTAN, strong X-ray limits need to be relaxed and dark matter overproduction needs to be avoided. We discuss postponing the dark matter production to lower temperatures, a reduced sterile neutrino contribution to dark matter, and a reduction of the branching ratio in photons and active neutrinos through cancellation with a new physics diagram. 
Both the Dodelson-Widrow and the Shi-Fuller mechanisms for sterile neutrino dark matter production are considered. 
As a final exotic example, potential consequences of CPT violation are discussed. 
\end{abstract}

\maketitle

\section{Introduction}
\label{sec:intro}
\noindent
The nature of dark matter (DM) is still an unsolved puzzle standing at the crossroads between cosmology, astrophysics and high energy physics.
From the perspective of the latter, the most popular and widely studied option for DM has been the Weakly Interacting Massive Particle (WIMP) \cite{Arcadi:2017kky}. Still, the mass window of viable DM candidates spans more than $50$ orders of magnitude, ranging from ultralight (``fuzzy") bosons \cite{Hu:2000ke} to primordial black holes \cite{Carr:2016drx}. In light of null-results for WIMP searches at colliders \cite{Kahlhoefer:2017dnp} and direct detection experiments \cite{Aprile:2018dbl}, this multitude of alternative possibilities is more motivated than ever.

In this paper we will focus on the fermionic DM at keV-scale, the so called ``keV sterile neutrinos"\footnote{In what follows, for brevity, we will refer to such particles simply as ``sterile neutrinos". We note that the sterile neutrinos at eV-scale are also well studied in the context of short-baseline neutrino oscillation anomalies \cite{Abazajian:2012ys}, but consideration of such states is beyond the scope of this work.}. This DM candidate was suggested already in the 90s in the pioneering paper by Dodelson and Widrow \cite{Dodelson:1993je}, who  proposed sterile neutrino production from active neutrinos through oscillations and collisions. Several years later, it was demonstrated that the production can be stimulated by resonant neutrino conversion \cite{Wolfenstein,Mikheev:1986gs,Mikheev:1986wj} in the presence of non-zero lepton asymmetries in the early Universe \cite{Shi:1998km}, the so-called Shi-Fuller mechanism. This mechanism was successfully embedded into the so-called $\nu$MSM \cite{Asaka:2005pn,Asaka:2005an} framework in which neutrino masses, DM and the baryon asymmetry of the Universe can be explained by employing only sub-TeV new physics\footnote{see also \cite{Baumholzer:2018sfb} where this is achieved in the framework of scotogenic model.}. 

Interest in sterile neutrino DM received further boost in 2014 when two groups independently observed an unidentified line at around $3.5$ keV in 
the X-ray spectra of galaxy clusters \cite{Bulbul:2014sua} and Andromeda \cite{Boyarsky:2014jta}. This discovery hinted to the exciting possibility that the line stems from the decay of 7 keV sterile neutrino.
While a DM explanation has been advocated also in more recent publications \cite{Cappelluti:2017ywp} it is fair to stress the existence of alternative explanations that do not involve beyond the Standard Model (BSM) physics; the most notable example is the charge-exchange mechanism \cite{Shah:2016efh}. The DM origin of the line was also criticized in \cite{Jeltema:2014qfa,Carlson:2014lla}.

The goal of this paper is not to resolve these astrophysical ``hints", but instead to focus on laboratory measurements. In other words, the main question we strive to answer is ``Can sterile neutrino DM be discovered in  terrestrial experiments and, if so, under which conditions?" The usual rule of thumb is that, given the present constraints from cosmology and astrophysics, the answer to the above question is negative. In particular, 
the mass and mixing values reachable by KATRIN correspond to a too large DM density if the simplest production mechanism is assumed. A rough estimate illustrating this tension is that 
in the most straightforward way of producing sterile neutrinos via oscillations and collisions in the plasma \cite{Dodelson:1993je}, the DM density can be written as \cite{Kusenko:2009up}
\begin{align} \label{eq:1}
\Omega_s h^2 \sim 0.12 \bigg(\frac{\sin^2 (2\theta)}{3.5\cdot 10^{-9}} \bigg) \bigg(\frac{m_s}{7 \, \rm keV} \bigg), 
\end{align}
where $m_s$ and $\theta$ are the sterile neutrino mass and mixing angle, respectively. 
This has to be compared to future experimental sensitivities around $\sin^2 (2\theta) \sim 10^{-6}$. 
Furthermore, the indicated parameter range 
is already strongly disfavored by structure formation and X-ray searches \cite{Merle:2015vzu}. In this paper we discuss several methods to reconcile these potential tensions. 

To be precise, in one of the considered scenarios we will assume that sterile neutrinos may not be the only part of the DM density in the Universe\footnote{Note that massive active neutrinos contribute to the DM density. Black holes contribute as well, though also with an unknown amount. Nevertheless, this demonstrates that a multi-component DM  scenario is definitely realized.}. Another option that we study is the relaxation of X-ray constraints by interference of known decay channel with another process involving additional new physics. One can also postpone the production of sterile neutrino dark  matter by starting their production at lower temperatures than in the standard case. We consider non-resonant and resonant production of $\nu_s$, and identify the parameter space that is accessible in future experiments.

While some of the discussed options may appear exotic, we regard such cases as an additional motivation for the experimentalists working on tritium beta decay and electron capture experiments to keep keV-scale DM searches in the focus of their BSM program. Tritium beta decay experiments are currently in the ``renaissance" era as the KATRIN experiment \cite{Franklin:2018adt,Mertens:2018vuu} is fully operative and has recently overriden the Mainz \cite{Kraus:2004zw} and Troitsk \cite{Abdurashitov:2015jha} measurements, improving the upper limit on neutrino mass by roughly a factor of two \cite{Aker:2019uuj}. In its next stage, KATRIN will be equipped with a novel detector system, TRISTAN \cite{Brunst:2019aod}, which will further improve the sensitivity on keV-scale sterile neutrinos by looking at the entire spectrum of the emitted electrons: in this condition, if a keV sterile neutrino with a mass $m_s < 18.6$ keV is produced, the corresponding signal will be a kink in the spectrum at an energy equal to the value of $m_s$. We wish to mention another 
tritium beta decay experiment, MATRIX \cite{MATRIX}, which is currently in a very early stage. The electron capture experiments such as ECHo \cite{Blaum:2013pfu}, HUNTER \cite{Smith:2016vku} and HOLMES \cite{Nucciotti:2018vyc} are also under way and they will provide complementary information. It is also worthwhile pointing out the Project 8 experiment \cite{Guigue:2017wzr} which will offer a novel detection technology in tritium beta decays by employing cyclotron radiation emission spectroscopy. Finally, PTOLEMY will be an ultimate experiment applying tritium beta decay \cite{Betti:2019ouf}\footnote{The main purpose of PTOLEMY will be to study cosmological relic neutrinos.}. Here our focus will be on the near term player KATRIN/TRISTAN and partly on ECHo. Since only the mixing with the electron flavor influences the beta decay probes, we fix the mixing with the other flavors to zero. Making connection with the usual WIMP language, the production of sterile neutrinos in beta decay or electron capture is the analogue of  collider searches. Indirect detection would be achieved by identifying their decay products, {\it i.e.\!}  X-rays. The direct detection analogue 
is probably a too difficult task as the cross section of non-relativistic sterile neutrinos in terrestrial detectors is tiny \cite{Ando:2010ye,Campos:2016gjh,Divari:2017coh,Li:2010vy}.\\

This paper is organized as follows. In \cref{sec:limits} we outline the current limits on sterile neutrino DM and present several ways to achieve their relaxation such that terrestrial experiments are sensitive to their mass and mixing parameters. In \cref{sec:DW} we apply these methods and focus on oscillational and collisional production of sterile neutrinos, with the goal to achieve a sufficiently large abundance of DM in the parameter space accessible 
in laboratory searches. In \cref{sec:SF} we scrutinize DM production in the presence of non-zero lepton asymmetries.  Finally, we summarize in \cref{sec:summary}.
\section{Current astrophysical and cosmological limits and their relaxation}
\label{sec:limits}

\subsection{X-ray limits}
\label{subsec:X-ray}
\noindent
The statistical limit that can be reached in the KATRIN/TRISTAN experiment is at the level of $\sin^2 (2\theta)\simeq 10^{-7}$ for $m_s$ in the ballpark of $10$ keV \cite{Mertens:2018vuu}. Here, $m_s$ and $\theta$ denote sterile neutrino mass and its mixing with \emph{electron} neutrinos, respectively. At the same time, limits from X-ray DM searches in this mass range are around $\sin^2 (2\theta)\simeq 10^{-11}$\,\cite{Perez:2016tcq} implying that in the vanilla scenario the detection of sterile neutrino DM in the laboratory is 
unachievable. 
In what follows, we discuss  general possibilities on how to relax X-ray limits. \\

\begin{figure}[t]
  \centering
   \begin{tabular}{ccc}
    \includegraphics[width=0.8\textwidth]{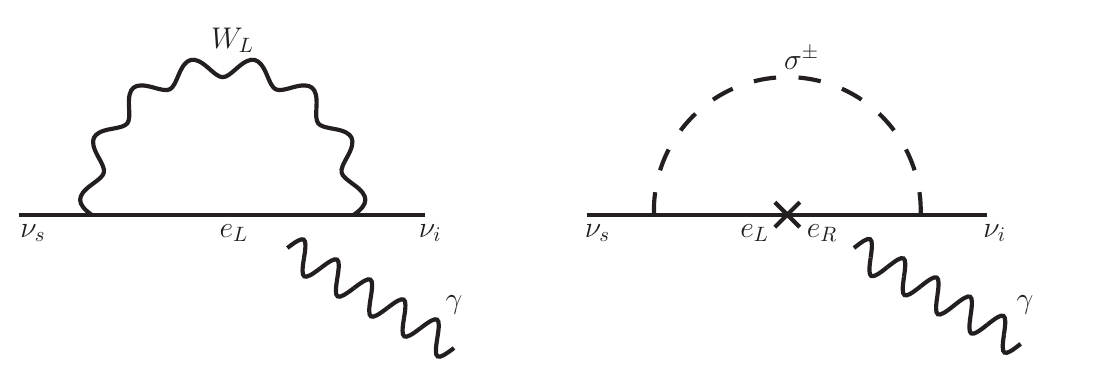}
     \end{tabular}
  \caption{Feynman diagrams for the decay of a sterile neutrino $\nu_s$ into an active neutrino and a photon. In each diagram, the photon line can be attached to any of the internal charged particles. In the left panel we show the ``standard" case with the electroweak $W_L$ boson, whereas in the right panel a new scalar particle is exchanged (see text for details).}
  \label{fig:1}
\end{figure}

$(i)$ \emph{DM cocktail}\\

\noindent
The flux of photons from $\nu_s$ decays is proportional to the number density of sterile neutrino DM as well as the decay rate $\Gamma_s$ into a photon and an active neutrino. It is given by 
\begin{align}
\Phi = \frac{\Gamma_s}{4\pi m_s} \int dl \, d\Omega \, \rho_{\text{s}}(l,\Omega)\,,
\label{eq:decay-rate}
\end{align}
where $l$ is the distance along the line of sight, $\Omega$ is the 
solid angle and $\rho_s$ equals the sterile neutrino DM energy density. 

The most straightforward way to relax the X-ray limits is to simply assume multi-component DM. That is, 
in addition to sterile neutrino DM there are other DM particles, such as WIMPs, that would make the total DM abundance, $\Omega_{\rm DM} h^2 = 0.12$ \cite{Aghanim:2018eyx}, consistent with observation and would not influence X-ray spectra. From the prefactor of \cref{eq:decay-rate} and the expression for the sterile neutrino decay rate via exchange of a SM $W_L$ boson \cite{Xing:2011,Pal:1981rm} (see the left diagram in \cref{fig:1}), 
\begin{align}
\Gamma_s= 1.38\cdot 10^{-32} \,\text{sec}^{-1}\,\left(\frac{\sin^2 (2\theta)}{10^{-10}} \right) \left(\frac{m_s}{1\,\text{keV}} \right)^5,
\label{eq:decay-classic}
\end{align}
one infers that at a given value of $m_s$, the present constraint on $\sin^2 (2\theta)$ is relaxed by a factor of
$\left(\Omega_{\rm DM} h^2/\Omega_s h^2\right)$, where $\Omega_s h^2$ is the $\nu_s$ abundance in the multi-component DM scenario. 

We denote this case here as \emph{cocktail} scenario. 
While the limit  clearly fades away as the abundance of $\nu_s$ gets reduced, it is not of our interest to consider scenarios in which the abundance of $\nu_s$ is greatly suppressed. Throughout this paper we will not show cases where the abundance of $\nu_s$ is less than 1\% of the total DM abundance.\\

$(ii)$ \emph{Reduced decay rate}\\

\noindent
From \cref{eq:decay-rate} it is clear that the reduction of the photon flux can also be obtained by diminishing the decay rate $\Gamma_s$. The general expression $\Gamma \propto \int d\text{Phase} \,|\mathcal{M}|^2$ decreases if the amplitude $\mathcal{M}$ is supplemented by another term which destructively interferes with the former, in addition to the expression corresponding to the left diagram in \cref{fig:1}. Before showing one particular example how this can be achieved within a minimal extension of the  Standard Model (SM), let us demonstrate how the relaxation of the limit scales in this case. To this end, we denote two amplitudes of interest with $\mathcal{M}_1$ and $\mathcal{M}_2$, where the latter one is introduced to partially cancel the former. If $\mathcal{M}_1$+$\mathcal{M}_2$ = $\chi \mathcal{M}_1$, where $\chi<1$, the total rate for the process decreases by a factor of $\chi^{-2}$. For instance, $\chi^2=1/10$ leads to an order of magnitude reduction of the decay rate. For a given value of $m_s$, the limit on  $\sin^2 (2\theta)$ gets relaxed by a factor of $\chi^{-2}$.

Now, let us illustrate this effect in the framework of a concrete minimal BSM realization\footnote{Recently, relaxation of X-ray limits was investigated in a supersymmetric framework \cite{Faber:2019mti}. The  authors found mild effects, due to diagrams involving heavy supersymmetric  particles and aditional mixing angle suppression.}. We adopt the ``cookbook" from Ref.\ \cite{Lavoura:2003xp}, where the author provides general formulae for 1-loop processes in which a fermion decays into a lighter fermion and a photon. 
The amplitude for $\nu_s (p_1)\to \nu_i(p_2) \gamma(q)$ ($i$ denotes an active neutrino in mass basis) is $\mathcal{M}= e \epsilon_\mu^* \xi^\mu$, where $\epsilon$ represents the polarization of the outgoing photon and \cite{Lavoura:2003xp}
\begin{align}
\xi^\mu= \bar{u}_{\nu_i} i \sigma^{\mu\nu} q_\nu (\sigma_L P_L + \sigma_R P_R) u_{\nu_s}\,.
\end{align}
Here, spinors are denoted with $u$, and $P_{L(R)}$ is left (right) projector.
For the left diagram in \cref{fig:1} we reproduced the results from the literature \cite{Bezrukov:2009th}:
\begin{align}
i\sigma_R = \frac{3}{64\pi^2}\frac{g^2}{m_{W_L}^2} m_s \,U_{ei}^* \sin\theta\,, 
\label{eq:usual_contribution}
\end{align}
Here $g$ and $m_{W_L}$ are the weak coupling constant and $W_L$ boson mass, respectively, $U_{ei}$ is an element in the first row of the leptonic mixing matrix and throughout the calculation we have consistently ignored sub-eV active neutrino masses.

To demonstrate that $\sigma_R$ in \cref{eq:usual_contribution} can be greatly reduced, we introduce a scalar doublet $\Sigma=(\sigma^0, \sigma^-)\sim (1,2,-1)$. The relevant part of the Lagrangian involving this state and $\nu_s$ reads
\begin{align}
\mathcal{L}\supset \lambda \,\bar{\nu}_s \Sigma^\dagger L_e + \lambda'\, \bar{e}_R \tilde{\Sigma}^\dagger L_e +h.c.\,,
\label{eq:lag}
\end{align}
where with $L_e$ we denote the lepton doublet of the first generation\footnote{For simplicity we assume $\Sigma$ interacts only with  leptons of the first generation.}. With these two Yukawa interactions one can construct the Feynman diagram given in the right panel of \cref{fig:1}, where the sterile neutrino decays via exchange of a charged particle from the $\Sigma$ doublet. 
Such interaction was already studied in the context of sterile neutrinos and the 3.5 keV line in \cite{Arcadi:2014dca}. By following again \cite{Lavoura:2003xp}, we find 
\begin{align}
i\sigma_R = \frac{\lambda \lambda'}{16\pi^2 m_\Sigma^2}m_e \left[\text{Log}\left(\frac{m_e^2}{m_\Sigma^2}\right)+1\right] U_{ei}^*\,, 
\label{eq:new_contribution}
\end{align}
where $m_e$ is the electron mass which appears in the amplitude due to the  chirality flip (see again the right diagram in \cref{fig:1}) that is necessary to preserve gauge invariance. Note that due to the Majorana nature of initial and final state fermions, one needs to include the complex conjugated process to the amplitude and this also yields non-zero $\sigma_L$. Still, already by using \cref{eq:usual_contribution,eq:new_contribution} one can obtain the condition for the complete cancellation between the amplitudes of the two considered diagrams: 
\begin{align}
\sin\theta=\left(\frac{-4 \lambda \lambda'}{3 g^2}\right) \frac{m_e}{m_s} \frac{m_{W_L}^2}{m_\Sigma^2} \left[\text{Log}\left(\frac{m_e^2}{m_\Sigma^2}\right)+1  \right].
\label{eq:cancel}
\end{align}
Taking $m_\Sigma \sim 1$ TeV and $\sin \theta\sim 10^{-4}$, which is in the ballpark of KATRIN/TRISTAN sensitivity, \cref{eq:cancel} yields to $\lambda \lambda' \simeq 10^{-6}$ for $m_s \sim 1$ keV. 

Let us estimate the size of the coupling $\lambda$ required to avoid the thermalization of $\nu_s$ with the SM bath. If $\nu_s$ thermalized, its abundance would overshoot the measured DM abundance by $1$-$2$ orders of magnitude \cite{Bezrukov:2009th,Heeck:2017xbu}. The first term in \cref{eq:lag} facilitates (inverse) decays $\sigma^\pm \leftrightarrow e^\pm + \nu_s$. The rate for this process should be smaller than the Hubble rate at $T\gtrsim m_\Sigma$ and this yields $\lambda \lesssim 10^{-7}$. Clearly, setting the coupling to such values does not allow sufficient relaxation of the $\nu_s$ decay rate as it would force the coupling $\lambda'$ to very large, practically unperturbative values. This suggests that \cref{eq:cancel} can be satisfied only with sub-TeV reheating temperature, a scenario in which these processes would be absent due to lack of energy to produce $\sigma^\pm$ after inflation. We should note that sub-TeV reheating is consistent with our central assumption, to be outlined in detail below, that the production mechanism for sterile neutrinos stems from active to sterile neutrino oscillations at $T \lesssim 100$ MeV. 

Finally, while \cref{eq:cancel} is the condition for the full cancellation (which corresponds to the complete absence of the X-ray signal), we will only require reduction of the decay rate that relaxes X-ray limits to a  level at which the $\nu_s$ parameters are accessible at terrestrial experiments, such as KATRIN/TRISTAN. While this still requires a certain fine-tuning of the parameters involved, it is a viable possibility. We would also like to emphasize that the $\Sigma$ doublet introduced in this section is not the only option for generating additional diagrams for $\nu_s \to \nu_i \gamma$, but only one model that we employed in order to demonstrate the effect. Several other scenarios that could be adopted in the context of keV-scale sterile neutrino DM are discussed in Ref.\ \cite{Arcadi:2014dca} (see also \cite{Arcadi:2013aba}).\\

$(iii)$ \emph{Decoupling beta decay from X-ray decay}\\

\noindent
Another way to evade X-ray limits and simplify sterile neutrino detection  is to decouple DM decay from beta decay. This has been demonstrated in Ref.\ \cite{Barry:2014ika} in a left-right symmetric framework. Here the beta decay could be mediated by exchange of a heavy $W_R$ boson, whose large mass $M_{W_R}$ suppresses the decay, in contrast to the small mixing angle $\theta$ in the standard scenario. The decay of the keV-scale sterile neutrino DM can occur as in the standard case via small mixing $\theta$, or via right-handed currents. The latter decay is suppressed due to the large $W_R$ mass and small mixing, as kinematically only decays in active SM neutrinos, which are sterile with respect to right-handed currents, are possible.  Hence, astrophysical observations only constrain $\theta$. This implies that beta decay can occur dominantly via $W_R$ exchange, which is the leading contribution as long as $(m_{W_L}/M_{W_R})^4$ is larger than $\theta^2$. While being an attractive possibility, this simple scenario is difficult to realize as LHC limits on $M_{W_R}$ are quite strong and $W_L$-$W_R$ mixing needs to be very small in order to suppress additional diagrams for DM decay (other options for new interactions of keV-scale neutrinos in beta decay are less constrained \cite{Ludl:2016ane}, but have not been studied yet regarding DM decay). The example demonstrates nevertheless that DM decay and beta decay may not necessarily be related, and therefore X-ray limits can be evaded. We will focus in the remainder of this paper on a one-to-one correspondence between DM and beta decay. 

\subsection{Structure formation limits}
\label{subsec:cosmo}    
\noindent 
The constraints from structure formation also play a role in reducing the available sterile neutrino parameter space. Typically, the strongest limits are obtained from Lyman-$\alpha$ forests \cite{Yeche:2017upn}, where for collisional production of $\nu_s$ the lower limit exceeds $20$ keV \cite{Yeche:2017upn} (somewhat milder constraints of $m_s\gtrsim 10$ keV arise in the case of resonant production due to a colder spectrum \cite{Schneider:2016uqi}).
However, it was pointed out that the gas dynamics of the inter-galactic medium can yield a striking effect in the absorption spectra \cite{Kulkarni:2015fga}, having consequently a potentially drastic impact on the sterile neutrino exclusion. To this end, Lyman-$\alpha$ bounds are often not shown in sterile neutrino literature \cite{Cherry:2017dwu,Perez:2016tcq} and we follow this conduct in our paper.  
We note here that even stronger limits on sterile neutrinos were recently derived from the 21-cm observation \cite{Schneider:2018xba}. However, in Ref.\ \cite{Boyarsky:2019fgp}, the authors concluded that such limit is unreliable due to  large star formation uncertainty at high redshifts.
This leads us to the constraints from Milky Way satellite counts \cite{Schneider:2016uqi} which were derived both for resonant and non-resonant production \cite{Merle:2015vzu,Cherry:2017dwu} (lower limits around $m_s\simeq 5-10$ keV). In the following sections we consider these two production mechanisms with certain modifications. Namely, in the remainder of the paper we chiefly employ low-temperature production of sterile neutrino DM for which structure formation limits have not yet been considered in the literature. Since the full derivation of this limit is beyond the scope of our work, we estimate the constraint from Milky Way satellite counts following a procedure given in \cite{Heeck:2017xbu}.
First, we calculate the averaged sterile neutrino momentum over temperature, $\langle p/T\rangle$, for the production mechanism under consideration (see for instance Eq.\ (4.3) in \cite{Bezrukov:2017ike} and the corresponding discussion on the impact of a temperature cutoff).
Then, we use  Eq.\ (7) of \cite{Heeck:2017xbu} to rescale the limit taken from \cite{Merle:2015vzu} that was derived for a thermal spectrum $\langle p/T\rangle\approx 3.15$, assuming the Milky Way mass of $3\times 10^{12} M_\odot/h$ \cite{Mwmass}. Such mass may be regarded as an upper value, given the multitude of previous measurements \cite{Wang:2015ala}; let us note that recent Gaia results seem to indicate a somewhat smaller Milky Way mass, but that measurement currently has large uncertainties \cite{Gaia}. We adopt a larger Milky Way mass in order to give the weakest allowed value for the constraint and this is in accord with our reasoning throughout the paper as already demonstrated for X-ray searches in \cref{subsec:X-ray}. 
The general dependence of the limit from Milky Way satellite counts on 
the Milky Way mass can be inferred from \cite{Merle:2015vzu}. Note that we will not show the structure formation limits in all of our plots; in cases where we discuss $\Omega_s h^2 < 0.12$ there may be another cold component which can drastically reduce the strength of the limit (the cocktail scenario discussed in \cref{subsec:X-ray}). This was also demonstrated in \cite{Merle:2015vzu}. 

Finally, constraints from satellite counts are stronger than those obtained from phase space considerations \cite{TG,Gorbunov:2008ka} and hence in our work we do not report the latter.

\subsection{Limits from Supernovae}
\label{subsec:sn} 
\noindent
Sterile neutrinos can be abundantly produced in a supernova core and carry away energy \cite{Shi:1993ee}. The observation of SN 1987A has led to  limits in the $m_s$-$\sin^2  (2\theta)$ parameter space \cite{Raffelt:2011nc,Arguelles:2016uwb,Suliga:2019bsq}, that can be competitive with  X-ray bounds.  Note, however, that these limits are derived for sterile neutrino mixing with either muon or tau neutrinos, whereas in this paper only mixing  with electron neutrinos is considered. For such a case, the constraints from supernovae are much weaker due to the following reason: if sterile neutrinos are mixed with electron neutrinos, there are typically two MSW  resonances that the neutrino encounters on its way out of the core. The appearance of two resonances stems from the charged current interaction 
term in the potential which is absent for muon and tau neutrinos \cite{Hidaka:2006sg,Hidaka:2007se}. With two resonances, the active neutrinos will typically convert to sterile species and then finally exit the core as active ones.  To this end, we do not further discuss limits from supernovae and would like only to emphasize the necessity for including sterile neutrino scenarios in forthcoming supernova simulations.

\section{Non-resonant production of sterile neutrinos and prospects for detection at terrestrial experiments}
\label{sec:DW}
\subsection{Production through Dodelson-Widrow Mechanism}
\label{subsec:DW-prod}
\noindent
First we consider the so-called Dodelson-Widrow \cite{Dodelson:1993je} production\footnote{Recently, this mechanism was also explored in the presence of neutrino self-interactions \cite{deGouvea:2019phk}.} in which sterile neutrinos are generated from active neutrinos through oscillations and scatterings in the plasma. 

The Boltzmann equation that governs the evolution of the sterile neutrino distribution function, $f_s(p,t)$, is \cite{Abazajian:2001nj}
\begin{align}
\frac{\partial}{\partial{t}}f_s(p,t) - H\,p\,\frac{\partial}{\partial{p}}f_s(p,t)\,
&\approx\,\frac{\Gamma_\alpha}{2} \langle P_m(\nu_\alpha \rightarrow \nu_s;p,t)\rangle \, f_\alpha(p,t)\,.
\label{boltzmann}
\end{align}
Here, $f_\alpha$ is the active neutrino distribution function, $H$ is the Hubble parameter and $\Gamma_\alpha \propto G_F^2\, p\, T^4$ is the interaction rate of active neutrinos with flavor $\alpha$ in the plasma.

The probability for oscillation into a sterile state is 
\begin{align}
\langle P_m(\nu_\alpha \to \nu_s; p,t)\rangle = \sin^2 (2\theta_M)\,\sin^2 \biggl(\frac{v\,t}{L}\biggr)\,,
\end{align}
where the square of the sine involving the sterile neutrino velocity 
$v$ simply averages to $1/2$ and 
 $\theta_M$ is the effective mixing angle in matter \cite{Abazajian:2001nj}
\begin{align}
\sin^2 (2\theta_M) = \frac{\Delta^2(p) \sin^2(2\theta)}{\Delta^2(p) \sin^2(2\theta) + D^2(p) + [\Delta(p) \cos(2\theta) - V_T(p) - V_L(p)]^2}\,.
\label{eq:matter}
\end{align} In \cref{eq:matter} several new quantities are introduced: $\Delta(p)\approx m_s^2/2p$, $D(p)=\Gamma_{e}(p)/2$ is the quantum-damping factor which represents the suppression of the production related to the loss of coherence due to the collisions of $\nu_e$ in the plasma. 
Furthermore, $V_T(p)$ is the thermal potential which equals \cite{Notzold:1987ik,Merle:2015vzu}
\begin{align}
V_T(p,T) = \pm \sqrt{2} \,G_F\, \frac{2\, \zeta(3) \, T^3}{\pi^2} \frac{\eta_B}{4} -\frac{8 \sqrt{2} \,G_F\, p}{3 m_Z^2} (\rho_{\nu_e} + \rho_{\bar{\nu}_e}) 
  -\frac{8 \sqrt{2}\, G_F\, p}{3 m_{W_L}^2} (\rho_{e^-} + \rho_{e^+})\,,
\label{thermalpot}
\end{align}
where the upper (lower) sign holds for neutrinos (antineutrinos), $\zeta(x)$ is the Riemann $\zeta$-function, $\eta_B = 6.05 \times 10^{-10}$ is the baryon asymmetry, $\rho_x$ denotes the energy density of species $x$,  $m_Z$ and $m_{W_L}$ are masses of weak gauge bosons and $G_F$ is the Fermi constant. 
Finally, $V_L$ is the potential related to the lepton asymmetry, which is vanishing for Dodelson-Widrow production. In \cref{sec:SF} we will scrutinize scenarios with non-zero lepton asymmetry which can in some cases dramatically influence $\sin^2 (2\theta_M)$. 

The Dodelson-Widrow production has a peak at $T\simeq 133 \left({m_s}/{\text{keV}}\right)^{1/3}$ MeV \cite{Abazajian:2001nj}. For solving \cref{boltzmann} and calculating the sterile neutrino DM abundance arising from this vanilla scenario (shown for instance in upper left panel of \cref{fig:cocktail}) we employ the publicly available $\mathtt{sterile-dm}$ code \cite{Venumadhav:2015pla} that $(i)$ robustly incorporates the effects of the QCD phase transition which occurs at temperatures where DM production peaks, and $(ii)$ appropriately treats the rapid change of relativistic degrees of freedom, $g_*$, in this temperature range. 

There is a caveat though; $\mathtt{sterile-dm}$ currently only contains muon neutrino interaction rates, while we are mostly interested in the electron flavor, given our focus on tritium beta decay experiments such as KATRIN, in which only electron neutrinos are produced. The differences between $\Gamma_\mu$ and $\Gamma_e$ are quantified in \cite{Merle:2015vzu,Asaka:2006nq} (see also \cite{Abazajian:2001nj}) and are below 10\% at temperatures around the QCD phase transition, leading to practically unobservable effects on logarithmic scales that we employ in our plots.   

At smaller temperatures, $T\lesssim 10$ MeV, the difference is larger, $\Gamma_e\approx 2 \Gamma_\mu$. While this is irrelevant for the vanilla Dodelson-Widrow production, it may yield more significant effects in scenarios where production starts at lower temperatures, which are of primary interest in this work (see \cref{subs,subsec:DW-results}). To this end, for production temperatures $T\lesssim$ 30 MeV we  solve \cref{boltzmann} with $\Gamma_e$ rates (taken from Ref.\ \cite{Merle:2015vzu}) in the regime $g_*=const$.    
After changing the integration variable from time to temperature ($dT/dt=-HT$) and employing Eq.\,(7) from \cite{Dodelson:1993je}, the LHS of \cref{boltzmann} can be expressed as $-HT \left(\partial f_s/\partial T\right)_{p/T}$. Now it is straightforward to perform an integration over $T$ at fixed $p/T\equiv r$ and obtain the following expression for the sterile neutrino distribution function:
\begin{align}
f_s(r)=\int_{T_\text{final}}^{T_\text{initial}} dT \frac{M_\text{Pl}}{1.66 \sqrt{g_*}\, T^3}\,\left[ \frac{1}{4} \frac{\Gamma_e(r,T) \sin^2 (2\theta) \left(\frac{m_s^2}{2 r T}\right)^2}{\left(\frac{m_s^2}{2 r T}\right)^2 \sin^2 (2\theta) +(\Gamma_e(r,T)/2)^2 +  \left(\frac{m_s^2}{2rT} \cos\theta - V_T(r,T) \right)^2} \right] \frac{1}{e^r +1}.
\label{eq:dist}
\end{align}
The corresponding expression for sterile antineutrinos, $\bar{f}_s(r)$, is obtained by simply changing appropriate signs in the potential (see \cref{thermalpot}). The lower integration boundary in \cref{eq:dist}, $T_\text{final}$, denotes a temperature of $\mathcal{O}(1)$ MeV at which active neutrinos decouple, while the upper one, $T_\text{initial}$, represents the starting temperature below which the production via oscillations is considered. The non-reduced Planck mass is denoted by $M_\text{Pl}$.

The DM relic abundance can then be obtained by performing an additional integral over $r$  \cite{Heeck:2017xbu}, 
\begin{align} 
\Omega_s h^2 = & \frac{s_0}{\rho_\text{crit}/h^2} \, m_s \,\frac{45}{4\pi^4 \, g_*} \int_0^\infty dr\, r^2 \bigg(f_s(r)+\bar{f}_s(r)\bigg)  \nonumber \\  =&\, 2.742 \times 10^2 \,\bigg(\frac{m_s}{\text{keV}}\bigg)  \,\frac{45}{4\pi^4 \, g_*} \int_0^\infty dr\, r^2 \bigg(f_s(r)+\bar{f}_s(r)\bigg),
\label{eq:relic}
\end{align}
where in the second equality we inserted numerical values for the entropy density at present times, $s_0$, and the critical energy density $\rho_\text{crit}$. 

\subsection{Critical Temperature}
\label{subs}
\noindent
There are two main prerequisites in order to have KATRIN/TRISTAN make potential discoveries of scenarios with keV-scale sterile neutrino DM: 
 First, the X-ray limits need to be sufficiently suppressed. This can be achieved and in \cref{sec:limits} we have provided several options. 
 Second, given successful suppression of the limits, the DM should not be overproduced for values of $m_s$ and $\sin^2 (2\theta)$ in the ballpark of KATRIN/TRISTAN sensitivity. The overproduction, however, does occur in the vanilla Dodelson-Widrow scenario (see Eq.\ (\ref{eq:1})) which features a peak production rate at around $T_\text{max}\simeq100$ MeV \cite{Dodelson:1993je}, as discussed already in \ref{subsec:DW-prod}.

Regarding the latter, one would achieve a suppression of the DM abundance if the production around $T_\text{max}$ can be somehow forbidden or at least heavily suppressed. We dub the temperature above which such suppression is achieved, {\it i.e.\!} the temperature at which DM production starts, as \emph{critical temperature} and denote it as $T_c$.
By appropriately adjusting the value of $T_c$ the desired amount of DM can be generated. Let us now briefly discuss the origin of $T_c$. 
 
 The obvious possibility is that it could simply be the reheating temperature, $T_c\equiv T_{\text{RH}}$. This  was already pointed out in the literature \cite{Yaguna:2007wi, Gelmini:2004ah}. It is fair to emphasize that low-reheating temperature scenarios have a flaw of not supporting any of the conventional mechanisms for the production of baryon asymmetry of the Universe, such as thermal leptogenesis \cite{Abazajian:2017tcc}; however alternative options, such as the Affleck-Dine mechanism, do exist \cite{Affleck:1984fy}. The lower limit on the reheating temperature stemming from Big Bang Nucleosynthesis (BBN) yields $T_{\text{RH}} \gtrsim 4-5$ MeV \cite{deSalas:2015glj,Hasegawa:2019jsa}. Note that for the model presented in Sec.\ \ref{subsec:X-ray} we also discussed an upper bound on $T_{\text{RH}}$ arising from the condition of avoiding thermalization of sterile neutrino DM; the requirement was that $T_{\text{RH}}$ needs to be smaller than the TeV-scale masses of new scalars in the $\Sigma$ doublet. 

Another possiblity is to arrange that above  $T_c$ the mixing in matter, $\sin^2 (2\theta_M)$, introduced in \cref{eq:matter}, is suppressed; this occurs if the sterile neutrino mass at $T>T_c$  either vanishes or is very large. The authors of Ref.\ \cite{Bezrukov:2017ike} presented several very appealing physical scenarios in which this is achievable. In particular, we would like to stress the realization in which the sterile neutrino mass is drastically increased in the early Universe through coupling with axion-like particles. Such particles can also play a role of a secondary DM candidate; this is potentially along the lines of the cocktail scenario that we discussed in \cref{sec:limits}. Note that in this case $T_c$ can attain values smaller than the lower bound for $T_\text{RH}$. 

After demonstrating these two general options for the suppression of DM production above certain temperatures, we turn now to the presentation of the corresponding phenomenological analysis.

\subsection{Results}
\label{subsec:DW-results}
\noindent
In what follows we show the results for the DM production and adopt two options for suppression of the X-ray limits, introduced in \cref{sec:limits}. Namely, we first focus on the DM cocktail case and then proceed by assuming the suppression of the sterile neutrino decay rate.
We will demonstrate in both of these cases how the interplay between successful suppression of astrophysical limits and the involvement of a critical temperature puts KATRIN/TRISTAN in a position to discover keV-scale sterile neutrino DM. As mentioned above, we focus here on the prospective measurements of KATRIN/TRISTAN. After the option to modify KATRIN in order to probe keV-scale neutrinos was proposed in Ref.\ \cite{Mertens:2014nha},  first preliminary measurements have already 
been made \cite{Brunst:2019aod}.  
We will use here the parameters from Ref.\ \cite{Mertens:2018vuu}.  
With the KATRIN experiment in its current settings an integral measurement with much reduced (by a factor $10^6$) source strength and 7 days of data taking, corresponding to $6 \times 10^{11}$ electrons, can be performed, resulting in the sensitivity denoted ``KATRIN sensitivity" in the plots to be presented. 
For better limits a new detector needs to be installed to cope with higher event rates, which is the basics of the planned TRISTAN project.  A differential measurement with 3 years of data taking, corresponding to $10^{16}$ electrons with a reduced source strength by a factor 100, results in a sensitivity to be denoted as ``TRISTAN sensitivity". Finally, the statistical sensitivity for 3 years of data taking with full source strength is denoted as ``TRISTAN statistical sensitivity". For comparison, we will also display the sensitivity of the ECHo experiment \cite{Blaum:2013pfu}, which will use electron capture (and hence probe neutrinos instead of antineutrinos in beta decay). The shown sensitivity corresponds to the anticipated final stage of the experiment when it will be possible to acquire around $10^{14}$ events in the $^{163}$Ho spectrum in a single year. The strongest sensitivity is achieved for $m_s\simeq 1$-$2$ keV and $\sin^2 (2\theta) \simeq 10^{-6}$ \cite{Adhikari:2016bei}.  \\

$(i)$ \emph{DM cocktail}\\

\begin{figure}[t]
  \centering
  \begin{tabular}{cc}
    \includegraphics[width=.48\textwidth]{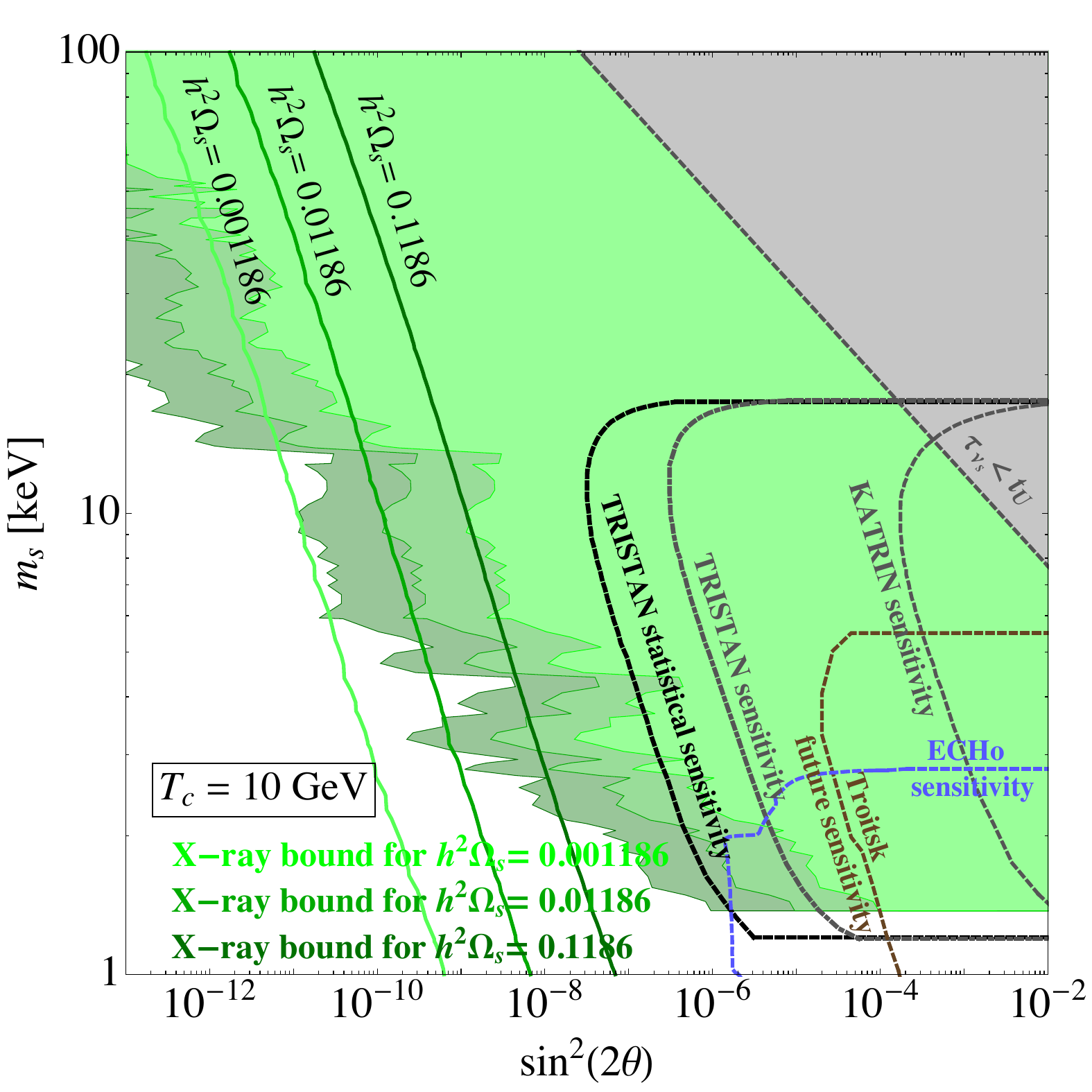} &
    \includegraphics[width=.48\textwidth]{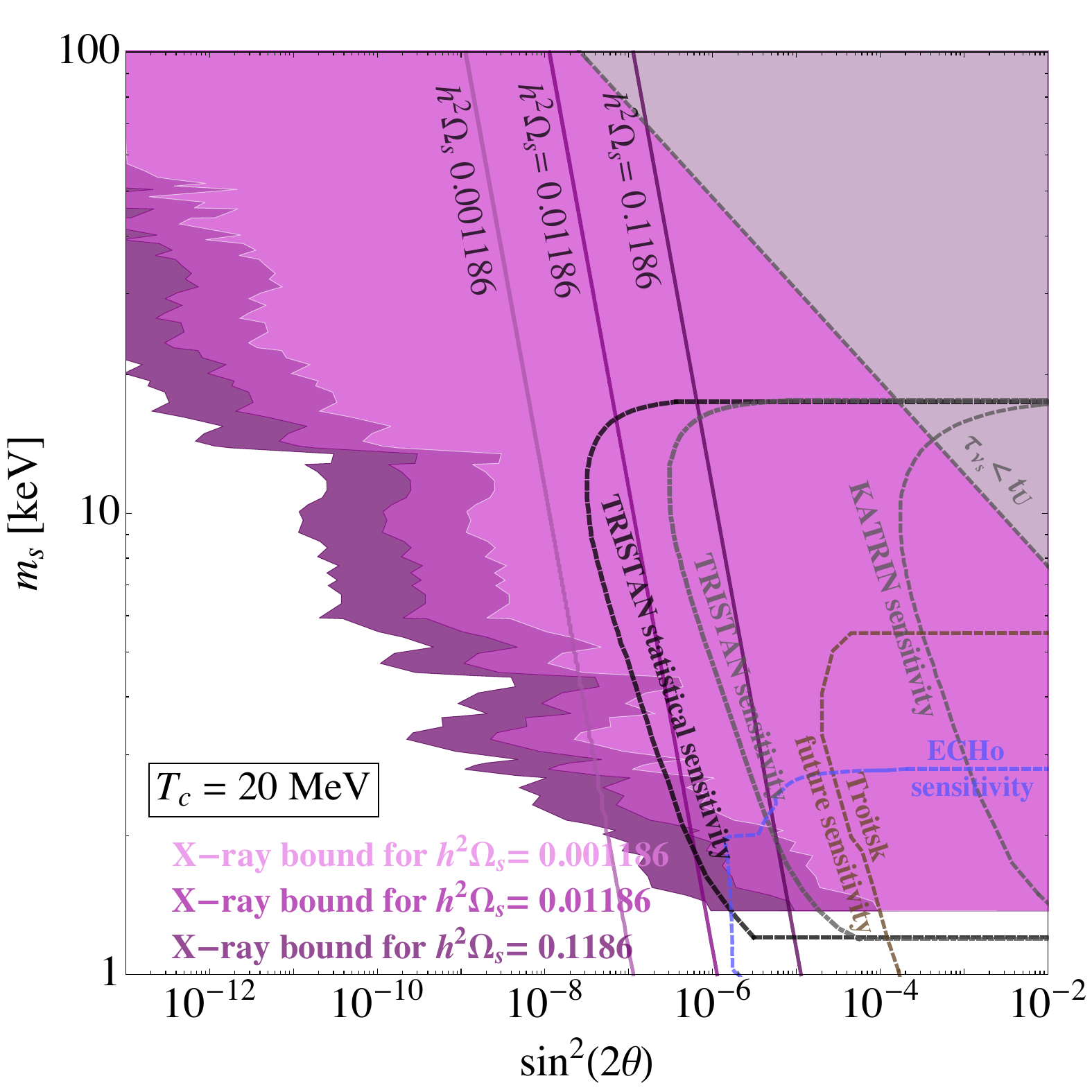} \\
    (a) & (b) \\[0.2cm]
    \includegraphics[width=.48\textwidth]{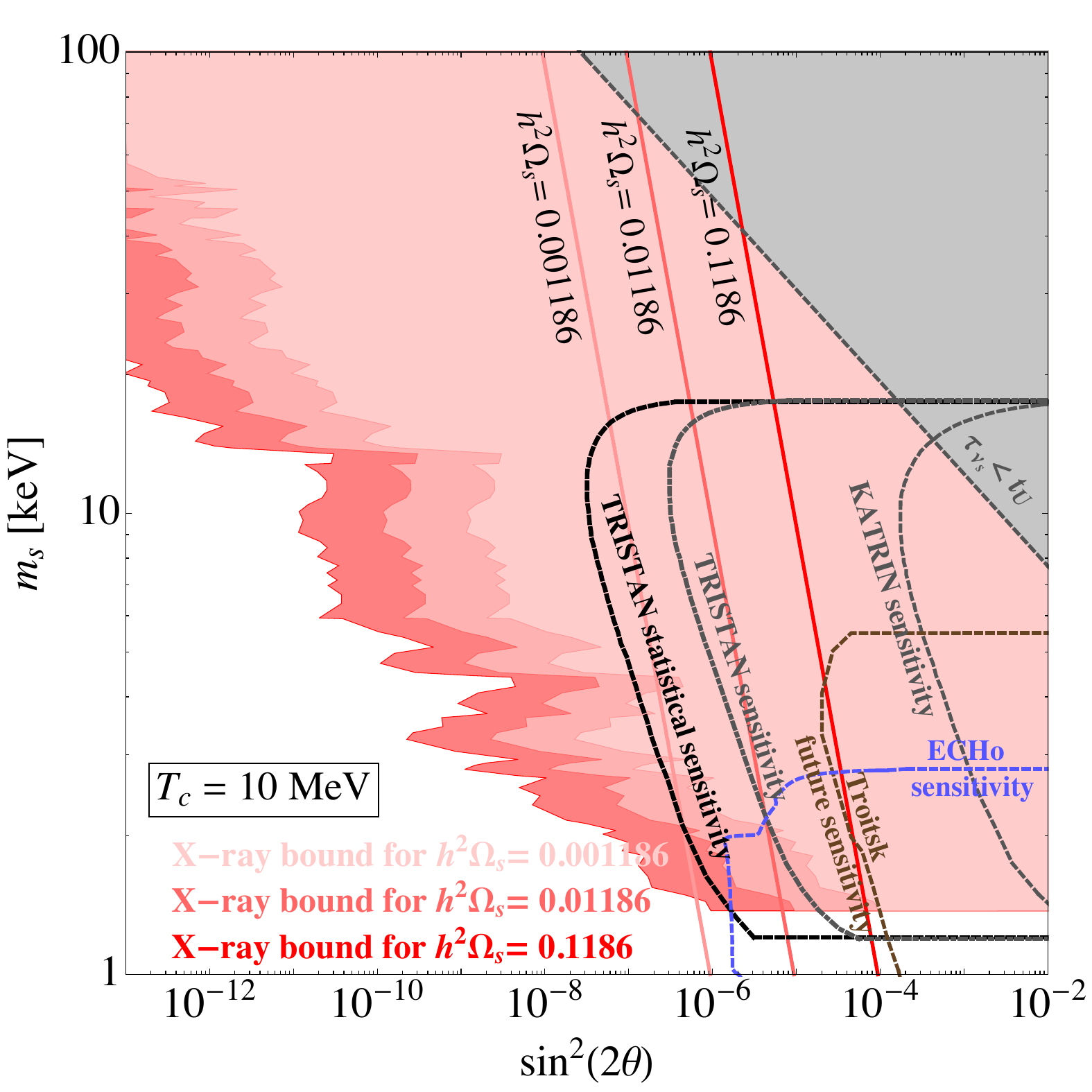} &
    \includegraphics[width=.48\textwidth]{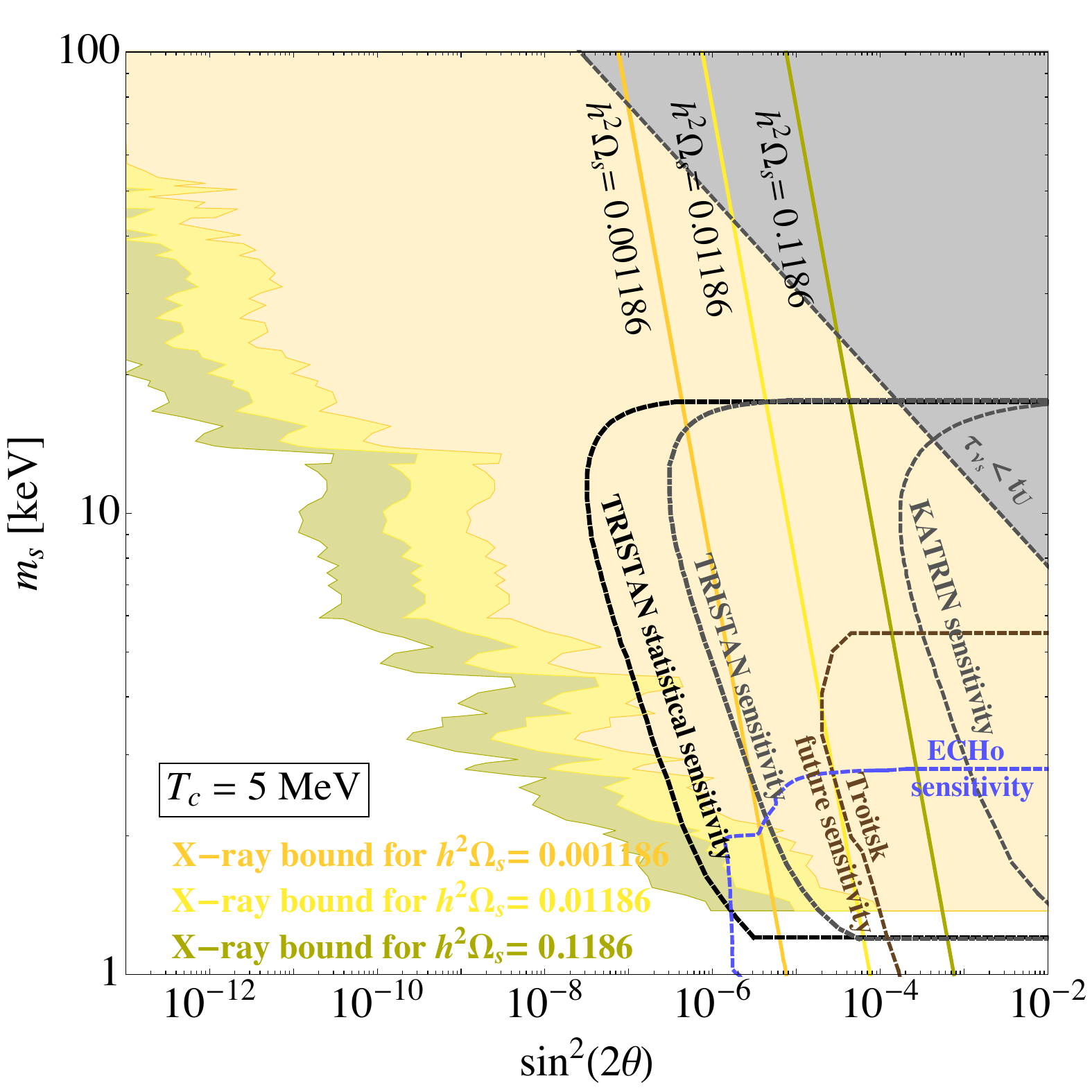} \\
    (c) & (d)
  \end{tabular}
  \caption{\emph{``Cocktail scenario"}: The colored regions in all panels show the X-ray constraints \cite{Horiuchi:2013noa,Perez:2016tcq,Ng:2019gch} for the indicated contributions of sterile neutrinos to the total DM density. In each panel these regions are shown in several shades of the same color. The lines, on the other hand, show the parameter space where the indicated amount of DM is produced via the Dodelson-Widrow mechanism below a critical temperature that is different (shown in black boxes) in each panel. We also show KATRIN/TRISTAN \cite{Mertens:2018vuu}, ECHo \cite{Blaum:2013pfu} and future Troitsk \cite{Abdurashitov:2015jha}
sensitivity for testing keV-scale sterile neutrinos as well as the constraint on sterile neutrino lifetime (gray-shaded region), arising from its main decay channel into three active neutrinos. By comparing X-ray constraints and respective lines for the DM production, and by focusing on the regions where KATRIN/TRISTAN is sensitive, one can infer that for $\Omega_s h^2=0.12$ and $\Omega_s h^2=0.012$ only very marginal parameter space is testable for $T_c \lesssim 20$ MeV. On the other hand, if sterile neutrinos represent 1\% of DM, KATRIN/TRISTAN is sensitive to $m_s\simeq$ $2$-$3$ keV, region in which X-ray limits are typically evaded; however in such case $T_c$ should not exceed $T_c\simeq$ 10 MeV.}
\label{fig:cocktail}
\end{figure}

\noindent
Let us first suppress the X-ray limits by reducing the amount of $\Omega_s h^2$ below $\Omega_{\rm DM} h^2 = 0.12$. 
The results for this approach are shown in \cref{fig:cocktail} for the following four values of the critical temperature: $T_c = 10\,\text{GeV}, 20\, \text{MeV}, 10\, \text{MeV}, 5\, \text{MeV}$. The first case 
clearly corresponds to the vanilla Dodelson-Widrow scenario, while in the remaining ones DM production is delayed to lower temperatures.
 In each panel, we compare the lines which indicate the regions in  parameter space where a certain amount of sterile neutrino DM is produced (we show cases in which sterile neutrinos account for 100\%, 10\% and 1\% of total DM) with accordingly modified X-ray constraints for those cases. For $T_c = 10\,\text{GeV}$, $\Omega_s h^2 = 0.12$ is in clear tension with X-rays in a vast portion of parameter space. The situation  improves upon assuming that sterile neutrinos represent only a subdominant component of DM. Still, the compatibility between DM production and the absence of X-ray limits occurs only in the parameter space that is far away from KATRIN/TRISTAN sensitivity. This expectedly changes upon lowering $T_c$ towards MeV-scale values. In particular, the effect of the critical temperature is evident in all remaining cases shown. The interesting region, in which KATRIN/TRISTAN will be sensitive, is intersected by $\Omega_s h^2 = 0.12$ curves; however, these regions are chiefly disfavored by X-ray limits that are still too strong in this case when sterile neutrinos would account for all the DM in the Universe. The only exception is for $m_s< \mathcal{O}(1)$ keV (see in particular the $T_c=20$ MeV and $T_c=10$ MeV panels) where DM production evades X-ray limits. Since KATRIN/TRISTAN sensitivity weakens in this mass range, only a rather marginal parameter space is testable; the situation is a bit better for the ECHo experiment which will probe somewhat smaller $m_s$ values.  The $\Omega_s h^2 = 0.1\times 0.12$ lines also intersect KATRIN/TRISTAN sensitivity curves 
but the X-ray limits still pose a problem, even upon relaxation.
 The viable range of masses increases however for $\Omega_s h^2 = 0.01\times 0.12$ to $m_s=2-3$ keV, but note that in this scenario KATRIN/TRISTAN is sensitive only for $T_c\lesssim$ 10 MeV.
In general, we conclude that KATRIN/TRISTAN could discover sterile neutrino DM in our  cocktail scenario provided $T_c$ is low enough. We wish to stress that we compared these results with those presented in Fig.\,3 of \cite{Bezrukov:2017ike} where the authors made similar claims. In particular, they also find that the consistency with X-ray limits can be achieved in the cocktail scenario. They also identified that, for $\Omega_s h^2 = 0.12$, the aforementioned $m_s\lesssim \mathcal{O}(1)$ keV range is viable. As stressed above, however, this is not particularly interesting for terrestrial experiments since the sensitivity in this mass range strongly declines. Let us finally note that the curves representing DM production were also compared to recent ones in the literature \cite{Gelmini:2019wfp,Gelmini:2019esj}, and agreement was found.\\

\begin{figure}[t]
\begin{center}
\includegraphics[width=.9\textwidth]{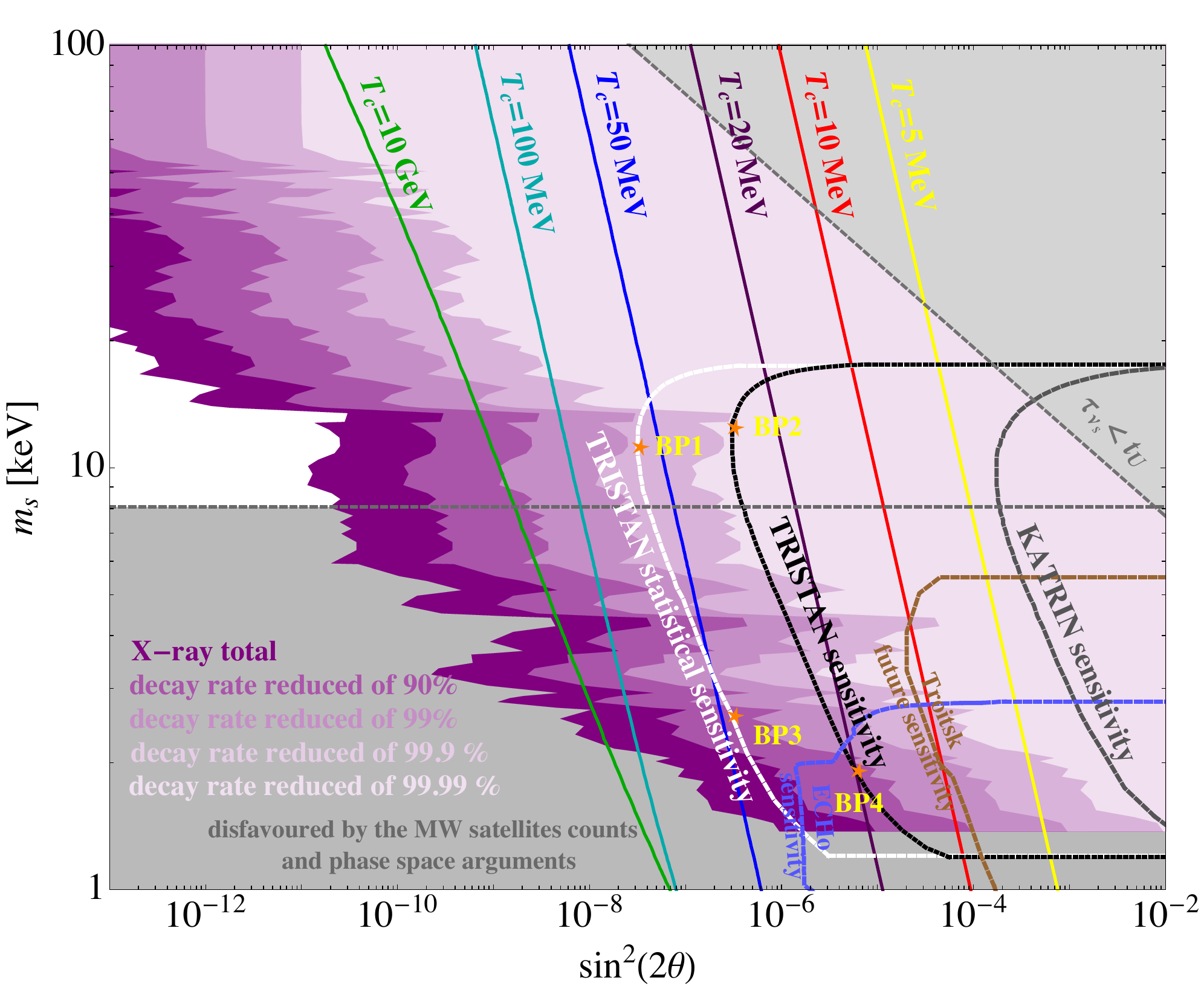}
\caption{\emph{X-ray suppression}: The shaded regions represent various X-ray limits, depending on the level of relaxation. We observe that the decay rate needs to be reduced by $3-4$ orders of magnitude in order to have  TRISTAN statistical sensitivity region free from X-ray limits.
The lines indicate the parameter space in which sterile neutrinos constitute the total DM abundance for the indicated values of $T_c$. We also show KATRIN sensitivity at different stages of the experiment (see previous figure) as well as the limit from structure formation (gray region). Stars indicate four benchmark points, which lie in the future sensitivity region of KATRIN/TRISTAN, for which we further explore connection between DM abundance and $T_c$ (see \cref{fig:benchmarks}).}
\label{fig:finalinverted}
\end{center}
\end{figure}

\begin{figure}[t]
\begin{center}
\includegraphics[width=.65\textwidth]{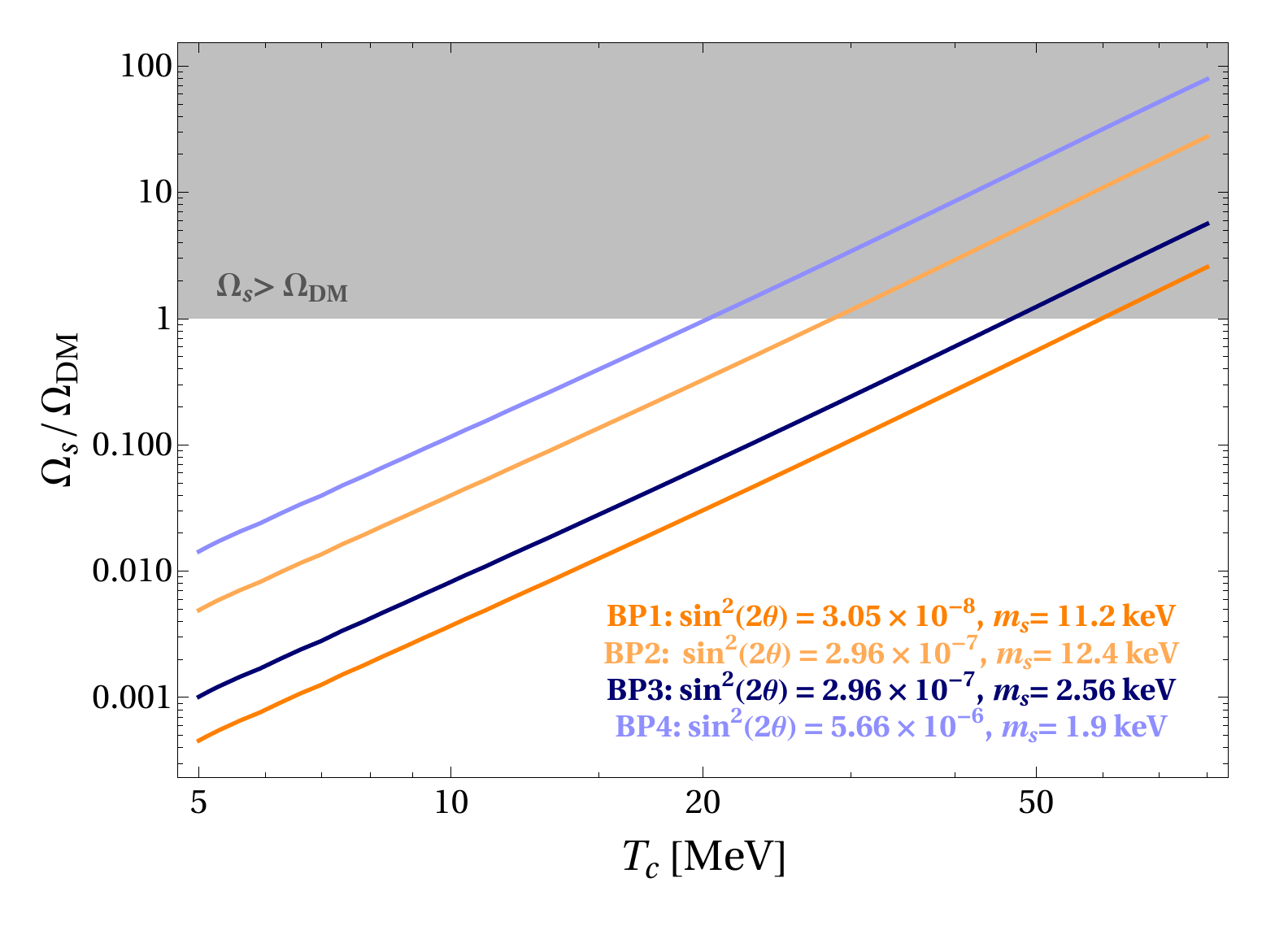}
\caption{The figure represents, for four chosen benchmark points, the dependence of the abundance of sterile neutrino DM (normalized to the observed amount of DM) on $T_c$. BP1 and BP3 lie on the TRISTAN statistical sensitivity curve, while BP2 and BP4 indicate the potential of TRISTAN to search for sterile neutrinos (see also \cref{fig:finalinverted}). In the considered low $T_c$ regime, the produced sterile neutrino DM abundance linearly depends on $T_c$ on our logarithmic scale. For our chosen benchmark points, $T_c$ values for which the correct amount of DM is produced lie between $T_c=20$ MeV and $T_c=60$ MeV.}
\label{fig:benchmarks}
\end{center}
\end{figure}

$(ii)$ \emph{Reduced Decay Rate}\\

\noindent
Now we consider the suppression of the X-ray limits due to the reduction of the decay rate of sterile neutrinos. In \cref{fig:finalinverted} we show the lines corresponding to different values of the critical temperature, $T_c$, and each of them represents the case in which sterile neutrinos account for all of the DM in the Universe. As already demonstrated in \cref{fig:cocktail}, from this figure one can also infer that when employing low $T_c$, lines are in the ballpark of
KATRIN/TRISTAN sensitivity. A partial relaxation of the X-ray bound is however still necessary in order to make such parameter space viable. In Sec.\  \ref{subsec:X-ray} we have discussed an example in which another diagram for the radiative decay of the sterile neutrino into an active neutrino  and a photon interferes destructively with the standard diagram. We show several relaxed X-ray bounds; in order to have a sufficient reduction of the X-ray limit such that the  statistical sensitivity region of  TRISTAN becomes viable, the sterile neutrino decay rate should be 3-4 orders of magnitude weaker. This is expected as KATRIN/TRISTAN sensitivity can reach $\sin^2 (2\theta)\sim 10^{-7}$, while present X-ray limits are in the ballpark of $10^{-11}$ for $m_s\simeq 10$ keV. 
Since here we assume that the total DM is constituted by sterile neutrinos, structure formation limits apply and to this end we show bounds from Milky Way satellite counts which disfavor detection at KATRIN/TRISTAN for $m_s\lesssim 8$ keV, leaving however still a significant portion of the parameter space open.\\

We discussed in this section two complementary options for relaxation of X-ray limits: the reduction of the decay rate as well as the cocktail scenario. If sufficient suppression is reached, the DM is still required to be produced in right amounts in the parameter space reachable by KATRIN/TRISTAN. Keeping the  Dodelson-Widrow mechanism in mind, we found that  a fairly low critical temperature is required in this case. Such values of $T_c$ can be well motivated in UV-complete realizations, for instance left-right symmetric models \cite{Bezrukov:2009th,Barry:2014ika}. Namely, we know that active neutrinos decouple at around 1 MeV. Sterile neutrinos, on the other hand, would decouple at somewhat larger temperature since the ``right-handed" gauge bosons are heavier than the electroweak ones. In order to avoid thermalization of sterile neutrinos (that would lead to overclosure of the Universe), one can opt for a low-reheating scenario in which $T_c=T_\text{RH}$ is smaller than the decoupling temperature. So, for a particular value of $T_c$, the lower bound on the mass of ``right-handed" gauge bosons can be set such that thermalization is avoided; while such limits are not competitive with the current LHC bounds, it is at least a demonstration that from eventual detection in KATRIN/TRISTAN we could also infer more about various scenarios. 
 
 Note that in \cref{fig:finalinverted} we also indicated four benchmark points, two of which lie at the border of TRISTAN sensitivity and the remaining two are on the line corresponding to the full statistical sensitivity of TRISTAN. For these points we illustrate in \cref{fig:benchmarks} the dependence of DM production on $T_c$. The figure clearly indicates that by increasing $T_c$ more DM gets produced; this statement is valid for low $T_c$. Once $T_c$ exceeds $T\simeq 100$ MeV, one is effectively in the vanilla Dodelson-Widrow regime and in this case the produced DM abundance is independent of $T_c$.

\section{Sterile neutrino production in the presence of lepton asymmetries and detection prospects}
\label{sec:SF}
\noindent
In this section we extend the discussion by asking the question whether DM whose production is influenced by the presence of non-zero lepton asymmetries in the early Universe can still be  detected at KATRIN/TRISTAN. Such asymmetries lead to the following potential \cite{Abazajian:2001nj}
\begin{align}
V_L = \frac{4 \sqrt{2} \,\zeta (3)}{\pi^2}\,G_{\rm F} T^3 \, L\,,
\label{vl}
\end{align}
where the lepton asymmetry is defined as $L = {(n_{\nu} - n_{\overline{\nu}})}/{n_\gamma}$, and we assume that it is non-vanishing  only in the active neutrino flavor that mixes with the sterile one; 
in our case, this is the electron flavor. Here, $n_{\nu}$, $n_{\overline{\nu}}$ and $n_\gamma$ are number densities of neutrinos, antineutrinos and photons, respectively. In the analysis, the term  in \cref{vl} should still of course be added to those given in \cref{thermalpot} that account for thermal properties. The usual effect stemming from the consideration of lepton asymmetries is the occurrence of resonant MSW transitions which can significantly increase the production in some regions of parameter space. Hence, in general, in order to produce DM in the amount that matches observations, smaller mixing angles in comparison to those in vanilla Dodelson-Widrow mechanism suffice \cite{Abazajian:2001nj}. While this would push successful regions in the parameter space even further from the domain of KATRIN sensitivity, one can again follow the strategy from \cref{sec:DW} and employ scenarios with low critical temperature that would delay and, more importantly, decrease production of DM. Hence, here we discuss the interplay between lepton asymmetries and low critical temperatures in the context of KATRIN. While we already know that considering low $T_c$ goes in the right direction (see \cref{subsec:DW-results}), the main purpose here is to quantify the magnitudes of lepton asymmetry for which the low $T_c$ regime is still testable at KATRIN/TRISTAN. Let us note that since neutrinos and antineutrinos have an opposite sign of the lepton asymmetry potential (\ref{vl}), resonant enhancement is only expected in one of the two species. Hence, in such scenarios, DM can be chiefly composed of either sterile neutrinos or antineutrinos\footnote{Throughout the paper we have abused notation by referring to sterile neutrino DM. We have done that since in the Dodelson-Widrow scenario, sterile neutrinos and antineutrinos are produced in equal amounts. However, we point out that in the resonant scenario the two components can be produced in vastly different amounts and hence it is required to precisely refer to neutrino or antineutrino.}. The sign of the asymmetry can be chosen in such a way that it is actually sterile antineutrinos that get produced as a dominant DM component; this is the species that is testable at KATRIN as in beta decays antineutrinos
are produced.

 We make use of $\mathtt{sterile-dm}$ that is currently the only available tool that can treat the evolution of asymmetries properly.
We have modified its source code in such a way that the production occurs only below a chosen $T_c$. We note, however, again that the code has only interaction rates for muon neutrinos implemented, and we are dominantly interested in the electron flavor; the error that this introduces is rather small. We also inferred that by making an additional check. Namely, we compared results for low $T_c$ in the absence of asymmetries obtained by $(i)$ running $\mathtt{sterile-dm}$ code and $(ii)$ solving \cref{eq:dist,eq:relic} numerically, and found very small differences in  the output of $\Omega_s h^2$. 

\begin{figure}[t]
\begin{center}
\includegraphics[width=.65\textwidth]{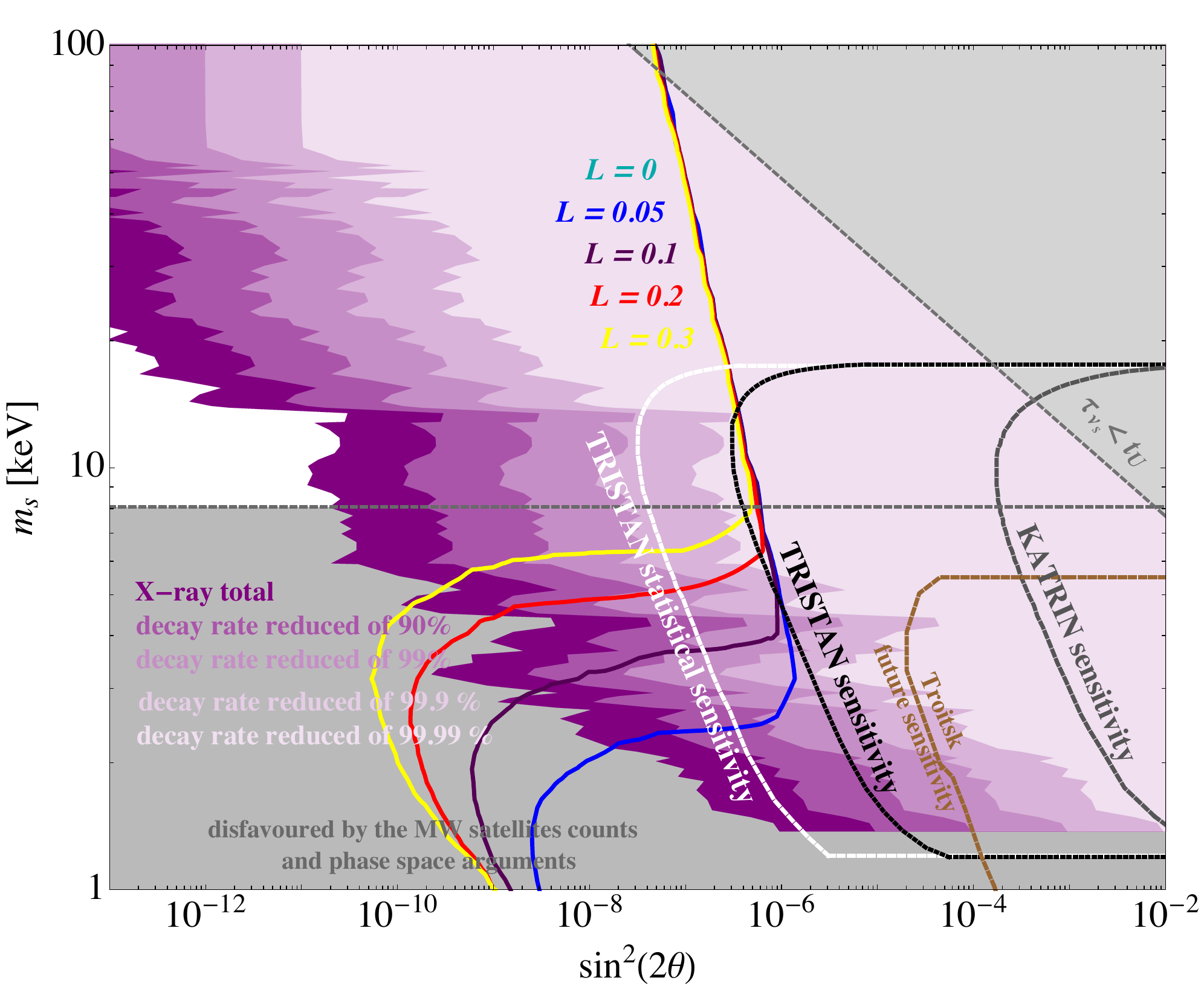}
\caption{\emph{Scenario with non-zero lepton asymmetries}: The curves indicate the parameter space with $\Omega_s h^2=0.12$ for $T_c=30$ MeV and several values of $L$ ranging between $0.05$ and $0.3$. These curves match the corresponding Dodelson-Widrow line at larger values of $m_s$ while resonant effects stemming from lepton asymmetries are prominent for  $m_s\lesssim 10$ keV. Part of the region in which the curves deviate from the  Dodelson-Widrow line is testable at KATRIN/TRISTAN, however in some tension with  structure formation limits. We also show relaxed X-ray limits as discussed in \cref{sec:DW} and presented in \cref{fig:finalinverted}.}
\label{fig:SF}
\end{center}
\end{figure}

The results of our analysis are shown in \cref{fig:SF} for a representative case of $T_c=30$ MeV. The initial asymmetry values  for which we show results are\footnote{Note that the final values of the asymmetry can  in some regions in  parameter space be at least a factor of a few times smaller in comparison to their initial values; this is  a consequence of resonant transitions.} $L=0.05$, $0.1$, $0.2$ and $0.3$. The chosen values are particularly large in comparison to the measured baryon asymmetry of the Universe. However, they are not excluded. Observational constraints on $L$ arise from CMB \cite{Oldengott:2017tzj,leptocmb} and BBN \cite{Castorina:2012md,Mangano:2011ip}; the reported limits between different analyses are conflicting and we will use values $L\lesssim 0.3$ that seem rather conservative. The curves in \cref{fig:SF} represent regions in which $\Omega_s h^2=0.12$ for the indicated values of the lepton asymmetry. We observe that for larger values of $m_s$ all curves with $L\neq 0$ match the Dodelson-Widrow ($L=0$) line for $T_c=30$ MeV. This is because at large $m_s$ the impact of $V_L$ is not competitive with  the  other terms in the potential, see Eqs.\ (\ref{thermalpot},\,\ref{vl}). At lower values of $m_s\lesssim 10$ keV, $V_L$ is much more relevant and we observe that the $\Omega_s h^2=0.12$ regions are shifted to smaller values of $\sin^2 (2\theta)$ due to resonances  that enhance the production. The regions with strongest resonant effects are not observable at KATRIN/TRISTAN. However, KATRIN/TRISTAN is still sensitive in the region where lines for $L\neq 0$ start departing from the Dodelson-Widrow line. If KATRIN/TRISTAN makes a discovery in such parameter space, a complementary cosmological determination of $T_c$ is required, as otherwise the discovery can be confused with the Dodelson-Widrow scenario employing a slightly larger $T_c$. We also note the apparent tension with satellite count limits in this region; the tension is milder for larger values of $L$.

\subsection{Sensitivity to CPT violation} 
\noindent
So far we discussed low $T_c$ scenarios for having the possibility to discover sterile neutrino DM at KATRIN/TRISTAN in the context of Dodelson-Widrow as well as resonant production. Here we will not consider $T_c$ but instead production starting at higher temperatures. As already stated, one of the species, either sterile neutrinos or antineutrinos, can be abundantly produced through the MSW effect. On the contrary, the production of the other species will be suppressed even with respect to the Dodelson-Widrow case because the sign of the potential is such that the effective mixing angle in matter receives suppression, becoming smaller than the vacuum value. Such suppression is exactly what is required to reach KATRIN/TRISTAN regions.  Note, however, that in order to fully exploit this possibility we need to kick sterile neutrinos out of the game and to this end we assume that sterile neutrinos \emph{are not mixed} with active ones. If that was not the case, we would simply be in the usual scenario with resonant production which is not interesting from the point of view of terrestrial experiments. The last assumption allows us to only focus on sterile antineutrinos that could have desired suppressed production stemming from a lepton asymmetry. Having different values of mixing angles for sterile antineutrinos and neutrinos implies CPT violation. It is  interesting to note that differences of mixing angles or masses are not very strongly constrained \cite{Barenboim:2017ewj} even for neutrinos and antineutrinos in the active sector. We wish to note that CPT violation was previously also considered in the context of eV-scale sterile neutrinos \cite{Barger:2003xm}.

The option of choosing different mixing angles for neutrinos and antineutrinos was not considered in $\mathtt{sterile-dm}$. So, if one executes the code for a given non-vanishing initial value of lepton asymmetry, as a consequence of resonant conversion of one of the species, 
the asymmetry can get reduced by an order of magnitude.
 This is not the physical picture appropriate for the tests of CPT violation as we need to take into account only sterile antineutrinos, with the sign of the potential chosen in such a way that they do not undergo resonant conversion, but instead feature suppression in the production. We do that by modifying again the $\mathtt{sterile-dm}$ source code; we suppress the usual lepton asymmetry and define a new one. We make the appropriate choice of the sign of the asymmetry for both species and in this way we arrange the situation in which both sterile neutrinos and antineutrinos receive suppression in the production stemming from the asymmetry term and there are no resonant effects. The full decoupling of either of the species is not possible in the code, but since in this realization both species are treated on equal footing, it is sufficient to simply divide the final DM abundance by a factor of two in order to access the DM abundance of sterile antineutrinos. The newly defined lepton asymmetry is set to remain constant throughout the evolution, as transitions between sterile and active antineutrinos are suppressed and therefore do not lead to significant changes of the initial value. 

\begin{figure}[t]
\begin{center}
\includegraphics[width=.65\textwidth]{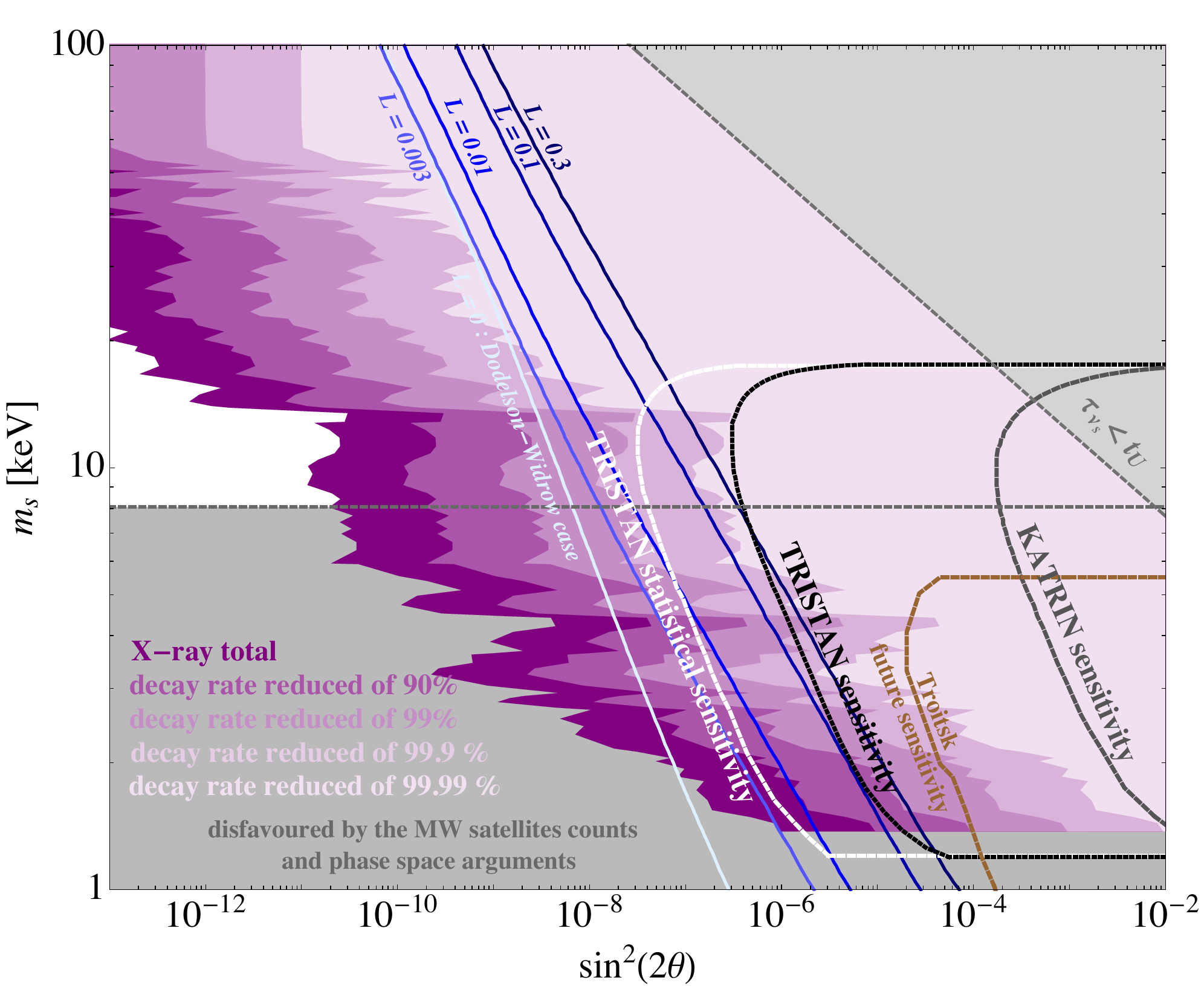}
\caption{\emph{CPT violating scenario}: The lines indicate 
values of lepton asymmetry $L$ for which sterile antineutrinos can generate an abundance of $\Omega_s h^2=0.12$. The discovery with TRISTAN is feasible only with $L\gtrsim 0.1$ while an order of magnitude smaller asymmetries are testable in the final KATRIN/TRISTAN stage. As in \cref{fig:finalinverted}, we show the relaxation of X-ray limits (for details see \cref{sec:limits,sec:DW}).}
\label{fig:CPTviolation}
\end{center}
\end{figure}

In \cref{fig:CPTviolation} we show the results of this analysis. All curves represent the case in which $\Omega_s h^2=0.12$ is produced in sterile antineutrinos, only  with different values of $L$. We observe that $L\simeq 0.01$ is enough for a discovery in the most sensitive stage of KATRIN/TRISTAN experiment. On the other hand, TRISTAN below the statistical sensitivity would be sensitive only if the asymmetry is very close to the upper limits from cosmology. The ultimate test of this possibility will be only possible if there is a synergy between beta decay and electron capture experiments. If there is discovery at KATRIN/TRISTAN, experiments such as ECHo and HUNTER should not  make an observation given their sensitivity only to sterile neutrinos (note that sufficient overlap in their mass sensitivity should exist).

\section{Summary and Conclusions}
\label{sec:summary}
\noindent
Laboratory production of DM particles would be a fundamental achievement of particle and astroparticle physics. For keV-scale sterile neutrinos, $\nu_s$,  this option may be possible with running/upcoming experiments such as KATRIN/TRISTAN. We have discussed the compatibility of a potential discovery of $\nu_s$ with their contribution to the DM density. Strong X-ray limits exist, and the parameters that are experimentally accessible would typically correspond to an overproduction of DM. However, we have presented ways to partially or fully avoid those constraints.

We have indicated the role of a critical temperature above which production of DM is forbidden or heavily suppressed. Working first in the framework of the Dodelson-Widrow mechanism, we achieved the desired suppression in the production by delaying DM production to $ \mathcal{O}(10)$ MeV temperatures. In addition, we found that if $\nu_s$ account for $1$\% of the total DM abundance, X-ray bounds are sufficiently relaxed in the region with $m_s\lesssim 3$ keV where KATRIN/TRISTAN is sensitive. Another possibility to avoid X-ray limits is to introduce new particles that could also mediate $\nu_s$ decays into active neutrinos and photons. By imposing destructive interference between contributing diagrams the decay rate would be reduced. We have given a concrete example of a model in which this could be achieved as well as have shown that the decay rate needs to be suppressed  by three to four orders of magnitude in order to make KATRIN/TRISTAN regions free from X-ray limits. In such case $\nu_s$ could account for the total DM density and can still be discovered in the laboratory. 

For completeness, we also considered DM production in the presence of non-vanishing lepton asymmetries, still assuming low critical temperatures of  $\mathcal{O}(10)$ MeV. Finally, we proposed a more exotic scenario which could be probed at KATRIN/TRISTAN if CPT is not an exact symmetry, namely if mixing angles of sterile neutrinos and sterile antineutrinos do not match.\\

In conclusion, there are ample options to produce one of the most elusive DM candidates in the lab, which would in addition provide exciting information on non-standard early Universe and particle physics.

\section*{Acknowledgments}
\noindent
We would like to thank John F.\ Cherry, Anton Chudaykin, Kenny Ng, 
Kathrin Valerius and Stefan Vogl for useful discussions. We are grateful to Johannes Herms for having pointed out an inaccuracy about the X-ray bound relaxation present in the previous version of the paper. CB acknowledges support by the IMPRS-PTFS and is also supported by DFG with grant RO 2516/5-1. WR is supported by the DFG with grant RO 2516/7-1 in the Heisenberg program. 
\bibliographystyle{JHEP}
\bibliography{refs}

\end{document}